\documentstyle[prl,aps,epsfig,floats,psfrag]{revtex}
\begin{document}
\draft
\newcommand{\be}{\begin{equation}}
\newcommand{\ee}{\end{equation}}
\newcommand{\bea}{\begin{eqnarray}}
\newcommand{\eea}{\end{eqnarray}}
\def\lsim{\raise0.3ex\hbox{$\;<$\kern-0.75em\raise-1.1ex\hbox{$\sim\;$}}}
\def\gsim{\raise0.3ex\hbox{$\;>$\kern-0.75em\raise-1.1ex\hbox{$\sim\;$}}}
\def\Frac#1#2{\frac{\displaystyle{#1}}{\displaystyle{#2}}}
\def\no{\nonumber\\}
\def\slash#1{\ooalign{\hfil/\hfil\crcr$#1$}}
\def\ep{\eta^{\prime}}
\def\susy{\mbox{\tiny SUSY}}
\def\sm{\mbox{\tiny SM}}
\textheight      250mm  

\twocolumn[\hsize\textwidth\columnwidth\hsize\csname@twocolumnfalse\endcsname
\rightline{\small IPPP/03/11 \, \, DCPT/03/22}
\vskip0.5pc 

\title{A possible supersymmetric solution to the discrepancy between $B\to \phi
K_S$ and $B\to \eta' K_S$ CP asymmetries}
\author{S. Khalil$^{1,2}$, and E. Kou$^1$}
\address{$^1$~IPPP, University of Durham, South Rd., Durham
DH1 3LE, U.K.\\
$^2$~Ain Shams University, Faculty of Science, Cairo, 11566, Egypt.\\}
\maketitle
\begin{abstract}
We present a possible supersymmetric solution to the discrepancy between 
the observed mixing CP asymmetries in $B\to \phi K_S$ and $B\to \ep K_S$. We show 
that due to the different parity in the final states of these processes, their 
supersymmetric contributions from the R-sector have an opposite sign, which naturally   
leads to $S_{\phi K_S} \neq S_{\ep K_S}$. 
We also consider the proposed mechanisms to solve the puzzle of the observed 
large branching ratio of  $B\to \ep K$ and their impact on $S_{\ep K_S}$. 
\end{abstract}
\vspace{-1cm}]
\noindent
Various measurements of  CP violation in B factory experiments 
have been opening a new era for the phenomenology of supersymmetric 
(SUSY) models. 
While in the standard model (SM), all the CP violating phenomena have to be explained by 
a single  phase in the Cabbibo-Kobayashi-Maskawa matrix, 
the SUSY models include additional new sources of  CP violation. 
Since these new effects can  manifest themselves in the CP asymmetries of various 
$B$-meson decays,  
the recent observed large discrepancy among the CP asymmetries of $B \to J/\psi K_S$, 
$B \to \phi K_S$ and $B \to \ep K_S$ have raised high expectations for 
indirectly unveiling  low energy SUSY \cite{Nir}. 

The measurement of the angle of the unitarity triangle $\beta(\phi_1)$ by the 
so-called golden mode $B \to J/\psi K_S$ \cite{jpsi1,jpsi2}: 
\begin{equation}
S_{J/\psi K_S}=\sin 2\beta(2\phi_1) =0.734\pm 0.054 
\end{equation}
is in a good agreement with the other measurements based on the standard 
model analysis. Flowingly, it has been shown that the effect from 
the SUSY particles in the box diagram which leads to the $B^0-\overline{B}^0$ mixing 
is typically small \cite{KG}. 
On the contrary, 
in summer 2002, the B factory experiments reported a surprising result for the measurement of 
$\beta$ by using the $B \to \phi K_S$ process. Since in the SM, 
$B \to J/\psi K_S$ and $B \to \phi K_S$ have the same $B^0-\overline{B}^0$ mixing part 
and do not have any additional CP violating phase in the decay process, 
the same value of $\sin 2\beta$ was expected to be extracted from them. 
Thus, the discovered large discrepancy \cite{phi1,jpsi2}
\begin{equation}
S_{\phi K_S}=-0.39\pm 0.41 \label{eq:2}
\end{equation}
has created quite a stir. Several efforts to explain this experimental data,  
in particular, by using SUSY models,  have been made.  
In Ref. \cite{KK}, it has been shown that this phenomena can
be understood without contradicting the smallness of the SUSY effect
on $B \to J/\psi K_S$ in the framework of the mass insertion approximation
which allows us to perform a model independent analysis of the SUSY
breakings \cite{HallRaby}. In this approximation, SUSY contributions are proportional to the
mass insertions $(\delta^d_{ij})_{AB}$ where $i,j$ and $A,B$ are the
generation and chirality indices, respectively.
While the measurement of  $B\to J/\psi K_S$ implies the smallness of 
$(\delta_{13}^d)_{AB}$, ($A,B=L,R$), the different generation mass insertion  
contributing to the $B\to \phi K_S$ process, $(\delta_{23}^d)_{AB}$,  
can deviate $S_{\phi K_S}$ from $S_{J/\psi K_S}$. 
In this letter, we discuss another measurement of $\sin 2\beta$ \cite{jpsi2,babar-etap} 
\begin{equation}
S_{\ep K_S}=0.33\pm 0.34 \label{eq:3}
\end{equation}
which has been thought to be problematic \cite{Nir}.  
Since $B\to \ep K_S$ gets  contributions 
from $(\delta_{23}^d)_{AB}$, $S_{\ep K_S}$ and $S_{\phi K_S}$ were expected 
to display similar discrepancy from $S_{J/\psi K_S}$.  
We will first show that 
although the magnitude of the SUSY contributions to these processes 
are indeed similar, $B \to \eta^{\prime} K_S$ has an opposite sign 
in the coefficient for the $RL$ and $RR$ mass insertions, which 
can naturally explain the experimental data. In fact, there is another open question on 
the $B \to \eta^{\prime} K$ process, the observed unexpectedly large branching ratio 
\cite{etabr1}. 
We will  further investigate the proposed new mechanisms to 
enhance the branching ratio of $B \to \eta^{\prime} K$ and their impacts on 
$S_{\eta^{\prime} K_S}$. 

The Effective Hamiltonian for the $\Delta B=1$ processes induced by gluino 
exchanges can be expressed as
\begin{equation}
H^{\small{\Delta B=1}}_{\mbox{\small eff}}=
-\frac{G_F}{\sqrt{2}}V_{tb}V_{ts}^*\sum_{i=3- 6, g}
\left[C_iO_i+\tilde{C}_i\tilde{O}_i\right]
\end{equation}
where  
the operators $\tilde{O}_{i}$ can be obtained from $O_i$ by exchanging $L\leftrightarrow R$. 
The Wilson coefficients 
$C_i$ and $\tilde{C}_i$ are proportional to $\delta_{LL, LR}$ and 
$\delta_{RR, RL}$, respectively. The definition of the operators and Wilson 
coefficients (and the effective Wilson coefficients below) can be found in \cite{KK}. Employing 
the naive factorisation approximation \cite{AG}, where all the 
colour factor $N$ is
assumed to be 3, the amplitude for the $B\to\phi K$ process  
can be expressed as \vspace{-0.3cm}
\begin{equation}
\overline{A}(\phi K)\!=\!-\frac{G_F}{\sqrt{2}}V_{tb}V_{ts}^*\sum_{i=3}^{6} 
\left[C_i^{\mbox{\small eff}}+\tilde{C}_i^{\mbox{\small eff}} \right]
\langle\phi\bar{K}^0 | O_i |\bar{B}^0\rangle \label{eq:6-00}
\end{equation}
where the matrix element is given by
\begin{equation}
\langle\phi\bar{K}^0 | O_i |\bar{B}^0\rangle = 
\left\{\frac{4}{3}X,\  \frac{4}{3}X,\  X,\  \frac{1}{3}X\right\}\ \ \  
(i=3 - 6)
\end{equation}
with $X=2F_1^{B\to K}(m_{\phi}^2)f_{\phi}m_{\phi}(p_K \cdot \epsilon_{\phi})$.  
$F_1^{B\to K}(m_{\phi}^2)=0.35$ GeV is the $B-K$ transition form factor and 
$f_{\phi}=0.233$ GeV is the decay constant of the $\phi$ meson. 
Since both $F_1^{B\to K}(m_{\phi}^2)$ and $f_{\phi}$ are insensitive to the 
chirality of the quarks, we used 
\begin{equation}
\langle\phi\bar{K}^0 | O_i |\bar{B}^0\rangle=\langle\phi\bar{K}^0 | \tilde{O}_i |\bar{B}^0\rangle
\label{eq:7-00}
\end{equation} 
to derive Eq. (\ref{eq:6-00}). 
On the other hand, the amplitude for $B \to \ep K$ can be written by: 
\begin{eqnarray}
\overline{A}(\ep K)&=&
 \frac{G_F}{\sqrt{2}}V_{ub}V_{us}^*\left[\sum_{i=1}^{2}C_i^{\mbox{\small eff}}\right]
\langle\ep\bar{K}^0 | O_i |\bar{B}^0\rangle \no
 &-&\frac{G_F}{\sqrt{2}}V_{tb}V_{ts}^*\left[\sum_{i=3}^{6}
\left(C_i^{\mbox{\small eff}}
- \tilde{C}_i^{\mbox{\small eff}}\right)\right]
\langle\ep\bar{K}^0 | O_i |\bar{B}^0\rangle .
\label{amplitude}
\end{eqnarray}
where we used
\begin{equation}
\langle\ep\bar{K}^0 | O_i |\bar{B}^0\rangle=-\langle\ep\bar{K}^0 | \tilde{O}_i |\bar{B}^0\rangle
\label{eq:9-00}
\end{equation} 
which is derived by the fact that the decay constant of $\ep$ is sensitive to the chirality 
of the quarks. The matrix element is given by:
\begin{eqnarray}
\langle\ep\bar{K}^0 | O_1 |\bar{B}^0\rangle &=& \frac{1}{3}X_2, \ \ \ \ 
\langle\ep\bar{K}^0 | O_2 |\bar{B}^0\rangle = X_2, \\
\langle\ep\bar{K}^0 | O_3 |\bar{B}^0\rangle &=& \frac{1}{3}X_1+2X_2+\frac{4}{3}X_3, \\
\langle\ep\bar{K}^0 | O_4 |\bar{B}^0\rangle &=& X_1+\frac{2}{3}X_2+\frac{4}{3}X_3, \\
\langle\ep\bar{K}^0 | O_5 |\bar{B}^0\rangle &=& \frac{R_1}{3}X_1-2X_2+(-1+\frac{R_2}{3})X_3, \\
\langle\ep\bar{K}^0 | O_6 |\bar{B}^0\rangle &=& R_1X_1-\frac{2}{3}X_2+(-\frac{1}{3}+R_2)X_3
\end{eqnarray}
with
\begin{eqnarray}
X1&=&-(m_B^2-m_{\ep}^2)F_1^{B\to\pi}(m_{K^{0}}^2)\frac{X_{\ep}}{\sqrt{2}}f_{K}, \nonumber \\
X2&=&-(m_B^2-m_{K^0}^2)F_1^{B\to K}(m_{\ep}^2)f_{\pi}\frac{X_{\ep}}{\sqrt{2}}, \nonumber \\
X3&=&-(m_B^2-m_{K^0}^2)F_1^{B\to K}(m_{\ep}^2)\sqrt{2f_{K}^2-f_{\pi}^2}Y_{\ep}, \nonumber \\
R1&=&\frac{2m_{K^0}^2}{(m_b-m_d)(m_s+m_d)}, \  
R2=\frac{2(2m_{K^0}^2-m_{\pi}^2)}{(m_b-m_s)(m_s+m_s)}\nonumber 
\end{eqnarray}
where $F_1^{B\to \pi}(q^2)=0.3$ GeV is the $B-\pi$ transition form factor and 
$f_{K(\pi )}=0.16(0.13)$ GeV 
is the decay constant of $K (\pi)$ meson. $q$ is the momentum transfer 
of the $b \to s$ transition.   
$X_{\ep}=0.57$ and $Y_{\ep}=0.82$, 
which correspond to $\theta_p=-20^{\circ}$, represent the 
rate of the $u\bar{u}+d\bar{d}$ and $s\bar{s}$ component in the $\ep$ \cite{R,K}. 
We use the following quark masses, $(m_d, m_s, m_b)=(0.0076, 0.122, 4.88)$ GeV.  
The tree contributions to $B \to \ep K_S$ is found to be less than 1\% 
and can be ignored. 

Numerical results on the ratio between SM and SUSY amplitudes 
for $m_{\tilde{g}}\simeq m_{\tilde{q}}=500~ \mathrm{GeV}$ are obtained as \cite{KK}
\bea
\left(\frac{A^{\susy}}{A^{\sm}}\right)_{\phi K_S} &\simeq&(0.23+0.04i)[(\delta_{LL}^d)_{23}
+(\delta_{RR}^d)_{23}]\no \ \ \ &\ &+(95+14i)[(\delta_{LR}^d)_{23}+(\delta_{RL}^d)_{23}] \label{aa1}\\
\left(\frac{A^{\susy}}{A^{\sm}}\right)_{\ep K_S} &\simeq& (0.23+0.04i)[(\delta_{LL}^d)_{23}
-(\delta_{RR}^d)_{23}]\no \ \ \ 
&\ &+(99+15i)[(\delta_{LR}^d)_{23}-(\delta_{RL}^d)_{23}].\label{aa2} 
\eea
where a parameter $q^2$ is chosen 
to be $m_b^2/4$. The variation of $q^2$ within the range of $m_b^2/6< q^2 <m_b^2/3$ 
causes $\pm 30$\% of theoretical uncertainty (see \cite{KK} 
for more detailed discussions on $q^2$ dependence). 
This problem of the unphysical $q^2$ dependence is recently solved by the new 
technology, the so-called QCD factorisation  (QCDF) approach \cite{BBNS}. 
Our result is consistent to the one using QCDF within the errors caused by 
the coefficients with the higher Gegenbauer terms \cite{footnote}. 
The small imaginary parts in Eqs. (\ref{aa1}) and (\ref{aa2}) are the strong phases  
which come from the QCD correction terms in the effective Wilson coefficient in Ref.\cite{AG}. 
The different sign for the contributions from the $O_i$ and $\tilde{O}_i$  
in $B \to \ep K_S$ (see Eq. (\ref{eq:9-00})) gives the minus sign for the coefficients of the 
$RL$ and $RR$ mass insertions in Eq. (\ref{aa2}). 
It is important to mention 
that this sign flip is not due to 
our using the naive factorisation approximation and 
would not be influenced by any other QCD corrections. 
Note that the absolute value of the mass insertion $(\delta_{AB}^d)_{23}$, which is relevant 
to the $b\to s$ transition is constrained by the experimental results for the branching 
ratio of the $B \to X_S \gamma$ decay: 
$\vert (\delta_{LL,RR}^d)_{23} \vert < 1$, $\vert(\delta_{LR,RL}^d)_{23} \vert \lsim 1.6 \times 10^{-2}$ \cite{bsgamma}.

We found that the coefficients for the each mass insertions are 
almost the same in $B \to \phi K$ and $B \to \ep K$ apart from the signs. 
Accordingly, we shall re-parametrise these ratios as: 
\begin{eqnarray}
\left(\frac{A^{\susy}}{A^{\sm}}\right)_{\phi K_S}\!&\!\equiv\!&\!R_{\phi }e^{i\delta_{12}}e^{i\theta_{\phi}}=
\delta_Le^{i\arg\delta_L}+\delta_Re^{i\arg\delta_R}, \label{eq:Rdef1}\\
\left(\frac{A^{\susy}}{A^{\sm}}\right)_{\ep K_S}\!&\!\equiv\!&\!R_{\ep}e^{i\delta_{12}}e^{i\theta_{\ep}}\simeq
\delta_Le^{i\arg\delta_L}-\delta_Re^{i\arg\delta_R} \label{eq:Rdef2}
\end{eqnarray}
where $\theta_{\phi(\ep)}$ and $\delta_{12}$ are CP violating and conserving phase 
differences between SM and SUSY, respectively. 
$\delta_{L}$ and $\delta_{R}$ include the contributions proportional to the mass insertions 
$(\delta_{LL,LR}^d)_{23}$ and $(\delta_{RR,RL}^d)_{23}$, respectively. 
Using these parameters, 
the mixing CP asymmetry is given as \cite{KK}
\begin{eqnarray}
S_{\phi K_S}\!&=&\! \Frac{\sin 2 \beta \!+\!2 R_{\phi} \cos \delta_{12} \sin(\theta_{\phi}\!+\!2 \beta)\!+\!R_{\phi}^2  
\sin (2 \theta_{\phi}\!+\!2 \beta)}{1+ 2 R_{\phi} \cos \delta_{12} \cos\theta_{\phi} +R_{\phi}^2} 
\nonumber\\
S_{\ep K_S}\!&=&\!\Frac{\sin 2 \beta \!+\!2 R_{\ep} \cos \delta_{12} \sin(\theta_{\ep}\!+\!2 \beta) \!+\!R_{\ep}^2  
\sin (2 \theta_{\ep}\!+\!2 \beta)}{1+ 2 R_{\ep} \cos \delta_{12} \cos\theta_{\ep} +R_{\ep}^2}  \nonumber
\end{eqnarray}
where we use $\sin 2\beta=0.73$ in our analysis. As can be seen from the above formulae, the 
strong phase enters only as $\cos\delta_{12}$ and the small strong phases 
found in Eqs. (\ref{aa1}) and (\ref{aa2}) lead to $\cos\delta_{12} =0.99$. 
Thus, we use $\cos\delta_{12}=1$ in the following.  

Here let us recall our main conclusions on $S_{\phi K_S}$ in Ref.~\cite{KK}.  
$S_{\phi K_S}$ as a function of $\theta_{\phi}$ behaves as a $\sin\theta_{\phi }$ curve 
taking the value $S_{\phi K_S}=0.73$ at the origin and bounded above by 1. 
A typical behaviour of $S_{\phi K_S}$ with 
$R_{\phi}=0.5$ and $\cos\delta_{12}=1$
is shown as the solid line in Fig.1. 
In the following, we will use this result as a reference and fix 
$R_{\phi}=0.5$ 
and also focus on  
the region $-3\pi/4\leq \theta_{\phi}\leq -\pi/2$ where $S_{\phi K_S}$ becomes  
negative.

Now let us discuss the $B \to \ep K_S$ process and see if we can 
explain out the puzzle of the observed mixing CP asymmetries: 
$S_{\phi K_S}\leq 0$ while $ S_{\ep K_S} \lsim S_{J/\psi K_S}$. 
First, we shall show our result without including the contributions from these new mechanisms 
suggested to enhance the branching ratio of $B \to \ep K$ in order to see explicitly  
the different behaviours of $S_{\phi K_S}$ and $S_{\ep K_S}$ due to the minus sign 
in Eq.(\ref{eq:Rdef2}). Having some possible SUSY models in our mind, we  perform 
a case-by-case study in the following. 

\begin{figure}[t]\vspace{-0.5cm}\begin{center}
\psfrag{phi}[l][l]{$S_{\phi K_S}^{\mbox{\tiny exp.}}$}
\psfrag{eta}[l][l]{$S_{\ep K_S}^{\mbox{\tiny exp.}}$}
\psfrag{s}[l][l]{$S_{\phi K_S, \ep K_S}$}
\psfrag{t}[l][l]{$\theta_{\phi}$}
\includegraphics[width=8cm]{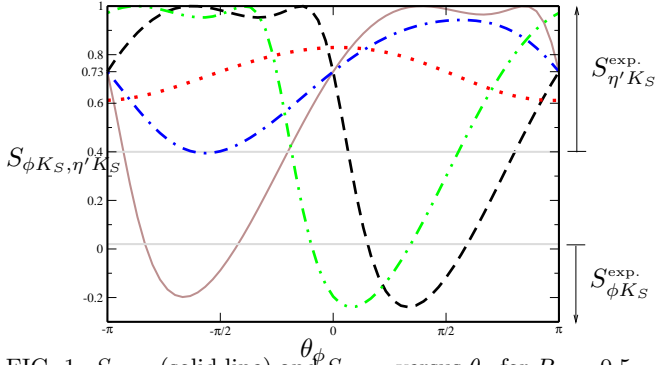}
\caption{$S_{\phi K_S}$ (solid line) and $S_{\ep K_S}$ versus $\theta_{\phi}$ for $R_{\phi}=0.5$ (dashed line for Case 1,
dotted line for Case 2, dash-dotted line for Case 3 and dash-double-dotted
for Case 4).}
\vspace{-0.5cm}
\end{center}\end{figure}

\vspace{0.1cm}
\noindent
{\bf Case 1) $|\delta_R|\gg|\delta_L|$} \\
Eqs. (\ref{eq:Rdef1}) and (\ref{eq:Rdef2}) lead to 
\begin{eqnarray}
R_{\phi}e^{i\theta_{\phi}}&=&|\delta_R|e^{i\arg\delta_R}, \\  R_{\ep}e^{i\theta_{\ep}}&=&|\delta_R|e^{i(\arg\delta_R+\pi)}. 
\end{eqnarray}
The CP asymmetry $S_{\ep K_S}$ as a function of $\arg\delta_R(=\theta_{\phi})$ is shown 
as a dashed line in Fig.1. $|\delta_R|$ is fixed to have $R_{\phi}=|\delta_R|=0.5$. 
As can be seen from this figure, $S_{\ep K_S}$ is always larger than the experimental 
data in Eq. (\ref{eq:3}) where $S_{\phi K_S}$ is within the experimental range. 
Note  that the $|\delta_L|$ dominated models give a same curve as $S_{\phi K_S}$.

\vspace{0.15cm}
\noindent
{\bf Case 2)} $|\delta_L|=|\delta_R|$ \\
In this case, Eqs. (\ref{eq:Rdef1}) and (\ref{eq:Rdef2}) are reduced to:  
\begin{eqnarray}
R_{\phi K_S}e^{i\theta_{\phi}}&=& 2|\delta_L|\cos\frac{\Delta\theta}{2}e^{i(\arg\delta_L+\arg\delta_R)/2}  \\
R_{\ep K_S}e^{i\theta_{\ep}}&=& 2|\delta_L|\sin\frac{\Delta\theta}{2}e^{i(\arg\delta_L+\arg\delta_R+\pi)/2} 
\label{eq:case2-2}
\end{eqnarray}
where $\Delta\theta=\arg\delta_L-\arg\delta_R$. 
We depict $S_{\ep K_S}$ as a function of 
$(\arg\delta_L+\arg\delta_R)/2(=\theta_{\phi})$ for $\Delta\theta=\pi/10$ as the 
dotted line in Fig.1.  
We fix $|\delta_L|$ so as to have $R_{\phi}=0.5$. The $\pi/2$ shift appearing in Eq.
(\ref{eq:case2-2}) can be clearly seen in the plot. 
It is also remarkable that in this case, not only the phase shift between 
$\theta_{\phi}$ and $\theta_{\ep}$ 
but also the amplitude difference which is given in terms of $\Delta\theta$ 
differentiate the behaviour of  $S_{\phi K_S}$ and $S_{\ep K_S}$. 
In particular, for small $\Delta\theta$, no matter what the value of 
$|\delta_L|$ is, $S_{\ep K_S}$ takes a value close to $\sin 2\beta$. 

\vspace{0.15cm}
\noindent
{\bf Case 3)} $\arg\delta_L=\arg\delta_R$\\
In this case,   we have  
\begin{eqnarray}
R_{\phi}e^{i\theta_{\phi}}&=&(|\delta_L|+|\delta_R|)e^{i\arg\delta_L} \\
R_{\ep}e^{i\theta_{\ep}}&=&\Delta|\delta|e^{i\arg\delta_L} 
\end{eqnarray}
where $\Delta|\delta|=|\delta_L|-|\delta_R|$. 
We show our results for $S_{\ep K_S}$ in terms of $\arg\delta_L(=\theta_{\phi})$ in
Fig.1 for  $R_{\phi}=|\delta_L|+|\delta_R|=0.5$ and 
$\Delta|\delta|=0.2$ (dash-dotted line).  
We found that the experimental bound gives a constraint of $0 \lsim \Delta|\delta|\lsim 0.4$.  


\vspace{0.15cm}
\noindent
{\bf Case 4)} $\arg\delta_R=\arg\delta_L+\pi/2$\\ 
In this case, we have 
\begin{eqnarray}
R_{\phi}e^{i\theta_{\phi}}&=&\sqrt{|\delta_L|^2+|\delta_R|^2}e^{i(\arg\delta_L+\alpha)} \\
R_{\ep}e^{i\theta_{\ep}}&=&\sqrt{|\delta_L|^2+|\delta_R|^2}e^{i(\arg\delta_L-\alpha)}  
\end{eqnarray}
where $\tan\alpha =|\delta_R|/|\delta_L|$. In Fig.1, we plot the result 
of $S_{\ep K_S}$  as  a function of $\arg\delta_L+\alpha(=\theta_{\phi})$ for 
$R_{\phi}=\sqrt{|\delta_L|^2+|\delta_R|^2}=0.5$ with $\alpha=5\pi/4$ 
(dash-double-dotted line). With the phase shift of $2\alpha$, one can 
have both $S_{\phi K_S}$ and $S_{\ep K_S}$ within their experimental range. 

We should comment that the above model independent analysis can be realised 
in well known SUSY models. 
For example, the SUSY models with Hermitian flavor structure which 
provide an interesting solution for the SUSY CP problem \cite{Hermitian}, 
have $\delta_{LR} = (\delta_{RL})^*$ with negligible $(\delta_{LL, RR})$, 
which is a realisation of Case 2 with $\Delta\theta=\pi$. 
Also in the SUSY seesaw models which 
are motivated by neutrino masses, $\delta_{RR}$ is much larger than $\delta_{LL}$ \cite{Murayama}.  
Therefore,  Case 1 can accommodate these models.

As mentioned, another large discrepancy is observed in the 
branching ratio of the $B \to \ep K$ process \cite{etabr1}: 
\begin{equation}
Br^{\mbox{\small exp.}}(B \to \ep K)=(55^{+19}_{-16}\pm 8 ) \times 10^{-6} 
\end{equation}
which is  2 to 5 times larger than the standard model calculation \cite{KS}. 
Since such a large deviation is observed only in the $B \to \ep K$ process, 
the new mechanisms based on the peculiarity of $\ep$ meson, for instance intrinsic charm 
\cite{CZ}
or gluonium contents of $\ep$ \cite{AS}, have been investigated. 
We shall discuss in the following: 
the SUSY effects to the branching ratio and 
the impacts of those new mechanisms 
on the mixing CP asymmetry $S_{\ep K_S}$. 

Let us first discuss the SUSY contributions 
and also the uncertainties from various 
SM parameters. In general, the SUSY contributions can be written as
\bea
Br(B\to \ep K) &=& Br_{SM}(B\to \ep K) \times \nonumber\\
&\ &\left[1+2\cos (\theta_{\ep}-\delta_{12})
R_{\ep}+R_{\ep}^2\right] \label{eq:br1}
\eea
Note that this equation can be applied to $B \to \phi K$ by replacing the indices. 
The input parameters which are used in our above analysis lead to 
$Br_{SM}(B\to \ep K)=13\times 10^{-6}$. In fact, this value is sensitive, especially to 
the $s$ quark mass and the value of $q^2$ in our calculation. For instance, 
$m_s=0.08$ GeV and $q^2=m_b^2/2$ give  
$Br_{SM}(B\to \ep K)=36\times 10^{-6}$. However, 
such a  small value of $m_s$ enhances the branching ratio of some similar 
processes like $B \to \pi K$ \cite{KS} and also a larger  $q^2$ is disfavoured by 
analysis of $S_{\phi K_S}$ \cite{KK}. 
A maximum enhancement from SUSY contributions can be obtained by  
$\theta_{\ep}=n\pi$, $n=0,1...$ and $\cos\delta_{12}=1$, which lead to
 $Br(B \to \ep K)=2.25\times Br_{SM}(B\to \ep K)$ for $R_{\ep}\simeq 0.5$. 
Interestingly our solution to reproduce the experimental result of 
$S_{\phi K_S}$ and $S_{\ep K_S}$ requires a shift between  
$\theta_{\phi}$ and $\theta_{\ep}$, which may 
suppress the leading SUSY contribution to the branching ratio for $B \to \phi K$. 
Thus, it is possible to enhance   $B\to \ep K$ 
without changing the prediction for $B\to \phi K$ too much. 
On the other hand, the other similar processes such as $B \to \pi K$ require more attention. 
Apart from its tree contributions, $B\to\pi K$ obtains as large SUSY contributions as 
$B \to \ep K$.
Therefore, we must not ignore the limitation given by these similar processes, 
which would be revealed as soon as more precise experimental data from those processes will be available. 

Now we turn to the new mechanisms proposed to enhance $Br(B\to \ep K)$ and its 
impacts on $S_{\ep K_S}$. 
We rewrite the amplitude in the following way, 
\be A(\ep K)=A^{\sm}_{\ep K_S}+A^{\susy}_{\ep K_S}+G^{\sm}+G^{\susy} \ee
where $G^{\sm}$ and $G^{\susy}$ are the new mechanism contributions to SM and SUSY, respectively. 
Flowingly, the branching ratio including the contributions from both SUSY and new mechanisms 
is modified to   
\bea
Br(B\to \ep K) &=& Br_{SM}(B\to \ep K) (1+r)^2\times \nonumber\\ 
&\ &\left[1+2\cos (\theta^{\prime}_{\ep}-\delta_{12})
R^{\prime}_{\ep}+R^{\prime2 }_{\ep}\right]. \label{eq:br2}
\eea
where $ r\equiv G^{\sm}/A^{\sm}_{\ep K_S}$ and 
$R^{\prime}_{\ep}e^{i\theta^{\prime}_{\ep}}=
(A^{\susy}_{\ep K_S}+G^{\susy})/(A^{\sm}_{\ep K_S}+G^{\sm})$.  
Note that $Br_{SM}(B\to \ep K)$ does not include the new mechanism contributions. 
Having the gluonium contributions in mind, 
we  parametrise the SUSY contributions from new mechanism as: 
$$
\frac{G^{\susy}}{G^{\sm}}=a\left[(\delta_{LL}^d)_{23} +(\delta_{RR}^d)_{23} \right] +
b\left[(\delta_{LR}^d)_{23} +(\delta_{RL}^d)_{23} \right]. 
$$
where $(\delta_{LL(LR)}^d)_{23}$ and $(\delta_{RR(RL)}^d)_{23}$ have a same coefficient 
due to the penguin process and also a same sign since the amplitude is proportional to only 
the $B-K$ transition form factor. 
Thus,  Eq. (\ref{aa2}) is modified to: 
\bea
R^{\prime}_{\ep} e^{i\theta^{\prime}_{\ep}}\!&\simeq\!&\! 
\left(\!\frac{0.23 \!+\!a\ r}{1\!+\!r}\! \right)(\delta_{LL}^d)_{23} 
+\left(\!\frac{101 \!+\! b\ r}{1\!+\!r}\!\right) (\delta_{LR}^d)_{23}\nonumber\\ 
\!&\!-\!&\! \left(\!\frac{101\!-\!b\ r}{1\!+\!r}\right)  (\delta_{RL}^d)_{23} 
\!-\! \left(\frac{0.23\! -\!a\ r}{1\!+\!r}\right)(\delta_{RR}^d)_{23} \label{eq:36}
\eea
Although the quantitative estimation of $r$ is difficult at the moment, the 
parameters $a$ and $b$ could be computed for a given mechanism. 
For the intrinsic charm contribution, we have $a=b=0$ since it come from a tree diagram. 
For the spectator gluonium contribution  
($G^{\susy}/G^{\sm}\simeq \sqrt{2}/(V_{tb}V_{ts}^*G_F)\ C_g^{\susy}/C_g^{\sm}$),   
we obtain $a=-1.2$ and $b=-585$ at the LL order. The spectator gluonium process means that 
the weak  $b\to s g$ transition (chromo-magnetic operator $O_g$) accompanied by 
one gluon emission from spectator is followed by two gluon fusion into gluonium in $\ep$ 
\cite{AKS,BN}.
Using these values in Eq. (\ref{eq:36}), we find that 
as $r$ increases $|\delta_L|_{\ep}$ is reduced and 
$|\delta_R|_{\ep}$ is enlarged. In fact, this does not disturb our previous explanation for   
the discrepancy between $S_{\phi K_S}$ and $S_{\ep K_S}$ especially because the sign 
in front of $|\delta_R|_{\phi}$ and $|\delta_R|_{\ep}$ remain different, which was a crucial point. 
In Fig. 2, we show the result for  the branching ratio  versus  $S_{\ep K_S}$  for 
Case $1-4$ including the spectator gluonium contribution. 
We fix $R_{\phi}=0.5$ and $\theta_{\phi}=-5\pi /8$ in order to have $S_{\phi K_S}\simeq -0.2$. 
As can be seen from this figure, we can have both of the CP asymmetry of $B\to \ep K_S$ and 
its branching ratio within the experimental limits in a significant  
range of SUSY parameters space.  
\begin{figure}[t]\vspace{-.8cm}\begin{center}\tiny
\psfrag{S}[l][l][1.5]{$ S_{\ep K_S}$}
\psfrag{B}[l][l][1.5]{$ Br(B \to\ep K)/Br^{\sm}(B \to\ep K)$}
\psfrag{r}[l][l]{$ r=0 \rightarrow 0.3$}
\psfrag{alpha}[l][l]{$\alpha=\pi $}
\psfrag{alpha2}[l][l]{$\rightarrow 5\pi/4$}
\psfrag{dt}[l][l]{$\Delta\theta=\pi/2 \rightarrow \pi/10$}
\psfrag{dd}[l][l]{$\Delta |\delta|=-0.5 $}
\psfrag{dd2}[l][l]{$\rightarrow 0.2$}
\psfrag{dd3}[l][l]{$\Delta |\delta|=0.1$} 
\psfrag{dd4}[l][l]{$\rightarrow 0.3$}
\psfrag{c1}[l][l][1.2]{Case 1}
\psfrag{c2}[l][l][1.2]{Case 2}
\psfrag{c3}[l][l][1.2]{Case 3}
\psfrag{c4}[l][l][1.2]{Case 4}
\psfrag{sexp}[l][l][1.2]{$S_{\ep K_S}^{\mbox{\tiny exp}}$}
\includegraphics[width=7.5cm]{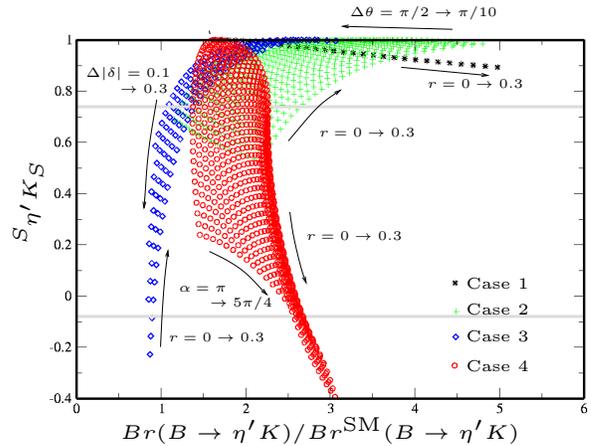}
\caption{A case-by-case study for  
the branching ratio of $B \to \ep K$ versus the mixing CP asymmetry 
$S_{\ep K_S}$. We assume that 
$R_{\phi}\simeq 0.5$ and $\theta_{\phi}\simeq -5\pi/8$ which lead to $S_{\phi K_S}\simeq -0.2$. 
The parameter $r$ represents the spectator gluonium contribution in SM. } 
\vspace{-0.6cm}
\end{center}\end{figure}

To conclude, we have considered possible supersymmetric contributions to 
the CP asymmetry $S_{\phi K_S}$ and $S_{\ep K_S}$. We showed that the discrepancy 
between their measurements can be naturally resolved by considering the different parity 
sensitivity of these processes to the SUSY contributions from R-sector   
that lead to $R_{\ep} < R_{\phi}$ and/or a phase shift 
between $\theta_{\ep}$ and $\theta_{\phi}$. We also studied the observed large 
branching ratio of $B\to \ep K$. 
We have considered the  new mechanisms proposed to enhance $Br(B\to \ep K)$ and 
their impacts on $S_{\phi K_S} - S_{\ep K_S}$ correlation.  
We have shown that a simultaneous solution for discrepancy between 
the CP asymmetries of $B \to \phi K_S$ and $B \to \ep K_S$ and the puzzle of 
the large branching ratio is possible for some SUSY models. 
More prices experimental data would allow us to 
draw more definite conclusions and shed light on compatible SUSY models with this 
solution.



\begin{thebibliography}{99}
\bibitem{footnote}
To include the chromo-magnetic annihilation diagram which would provide a 
significant contribution in the SUSY analysis, we need QCDF results at one  order higher 
in $\alpha_s$. In order to avoid a misleading result, we did not 
include any annihilation contributions in our QCDF estimate.
\end{thebibliography}
\end{document}